\documentclass[twocolumn,jcp,aip,superscriptaddress,showpacs,citeautoscript]{revtex4-1}

\usepackage{graphicx}
\usepackage{dcolumn}
\usepackage{bm}
\usepackage{flafter}
\usepackage{float}
\usepackage{lettrine}
\usepackage{amsmath}
\usepackage{amssymb}
\usepackage{color}
\usepackage{epsfig}
\graphicspath{{eps+pdf/}}

\begin{document}

\title{Ethane and methane at high pressures: structure and stability}

\date{\today}

\author{Elissaios Stavrou}
\email{elissaios.stavrou@gtiit.edu.cn}
\affiliation {Earth and Planets Laboratory, Carnegie Institution of Washington, Washington, D.C., USA}
\affiliation{Materials and Engineering Science Program, Guangdong Technion-Israel Institute of Technology, Shantou, Guangdong, 515063, China}
\affiliation{Technion-Israel Institute of Technology, Haifa, 32000, Israel}

\author{Alexander A. Maryewski}
\affiliation{Skolkovo Institute of Science and Technology, 3 Nobel St., Moscow 143026, Russian Federation.}

\author{Sergey Lobanov}
\affiliation{Earth and Planets Laboratory, Carnegie Institution of Washington, Washington, D.C., USA}
\affiliation{GFZ German Research Center for Geosciences, Section 4.3, Telegrafenberg, 14473 Potsdam, Germany.}

\author{Artem R. Oganov}
\affiliation{Skolkovo Institute of Science and Technology, 3 Nobel St., Moscow 143026, Russian Federation.}

\author{Zuzana Kon\^{o}pkov\'{a}}
\affiliation{DESY Photon Science, D-22607 Hamburg, Germany}
\affiliation{European XFEL GmbH, Holzkoppel 4, D-22869 Schenefeld, Germany}

\author{Vitali B. Prakapenka}
\affiliation{Center for Advanced Radiation Sources, University of Chicago, Chicago, IL, United States.}

\author{Alexander F. Goncharov}
\email{agoncharov@carnegiescience.edu}
\affiliation{Earth and Planets Laboratory, Carnegie Institution of Washington, Washington, D.C., USA}

\begin{abstract}
We have performed a combined experimental and theoretical study of ethane and methane at high pressures up to 120 GPa at 300 K using x-ray diffraction and Raman spectroscopy and the USPEX  ab-initio evolutionary structural search algorithm, respectively. For ethane, we have determined the crystallization point, for room temperature, at 2.7 GPa and also the low pressure crystal structure (Phase A). This crystal structure is orientationally disordered (plastic phase) and deviates from the known crystal structures for ethane at low temperatures. Moreover, a pressure induced phase transition has been identified, for the first time, at 13.6 GPa to a monoclinic phase B, the structure of which is solved based on a good agreement of the experimental results and theoretical predictions. For methane, our x-ray diffraction (XRD) measurements are in agreement with the previously reported high-pressure structures and equation of state (EOS).  We have determined the EOSs of ethane and methane, which provides a solid basis for the discussion of their relative stability at high pressures.
\end{abstract}

\maketitle

\section{Introduction}
Methane is one of the most abundant hydrocarbon molecules in the universe and is expected to be a significant part of the icy giant planets (Uranus and Neptune) and their satellites \cite{Hubbard1981,Podolak1995}. Ethane is one of the most predictable products of chemical reactivity of methane at extreme pressures and temperatures \cite{Ancilotto1997,Kolesnikov2009,Gao2010, Lobanov2013} . Moreover, the broad range of thermodynamic conditions at which hydrocarbons are present in the universe (from below 100 K to 10000 K and pressures up to 1 TPa) determines the importance of understanding the physics and chemistry of hydrocarbons at extreme pressure and temperature. Both methane and ethane have been found in the planetary atmospheres \cite{Moses1995}.

In spite of numerous experimental and theoretical studies \cite{Hirai2008,Casely2010,Casely2014,Umemoto2002,Sun2009,Chen2011,Zhu2012,Proctor2017}, the structure and relative stability of methane and other hydrocarbons at high pressures, even at room temperature, remains controversial even at moderate pressures. At room temperature, methane solidifies in a plastic fcc phase (orientationally disordered) that is  stable up to 5.4 GPa \cite{Hazen1980}. At higher pressures methane adopts orientationally disordered (based on the Raman spectroscopy data which show splitting of the C-H stretch modes \cite{Hirai2008,Chen2011}) phases: a) a rhombohedral structure (phase A)   up to 9 GPa and b) a cubic structure (phase B) up to 25 GPa  \cite{Casely2010,Casely2014}. The hydrogen positions could have been determined only for phase A (below 9 GPa). A phase must be disordered based of NO splitting of $\nu_1$ and $\nu_3$. Ref. \onlinecite{Casely2010} (neutron and X-ray diffraction measurements) suggest that it may be partially ordered. The transition between phase A and phase B is very sluggish; if pressure is increased quickly, yet another phase named pre-B is formed, which shows Raman spectra similar to phase A and XRD patterns similar to phase B \cite{Hirai2008}. The exact structure of the high-pressure (HP) phase above 25 GPa is unknown. It was proposed that methane would behave as a \textquotedblleft bad \textquotedblright noble gas and assume an hcp structure at high pressure \cite{Bini1997}. However, the structure of methane can be well indexed by a cubic phase up to 202 GPa \cite{Sun2009}. In the later study two distinct cubic phases, namely a simple cubic (SC) stable up to 94 GPa and high-pressure cubic (HP-C), were reported.

In contrast to rather rich data on methane, very little is known about ethane at high pressure \cite{Kolesnikov2009,Lobanov2013,Kurnosov2006,Podsiadlo2017}. Theoretical structure search, which includes a possible composition change between various hydrocarbons,  suggest that methane is stable to almost 100 GPa \cite{Gao2010}, while ethane is more stable at higher pressures where methane disproportionates  forming also butane, methane-hydrogen compounds (CH$_4$)$_4$(H$_2$)$_2$ and (CH$_4$)$_2$(H$_2$)$_3$ (similar to those reported at lower pressures \cite{Somayazulu1996}) and molecular H$_2$ \cite{Naumova2019}.
Above 300 GPa the only stable phases are H$_2$ and diamond in a qualitative agreement with the earlier work \cite{Ancilotto1997}. All these phases can be metastable in a wide pressure range and the kinetic hindrance can be overcome via a high temperature treatment. However, nominally metastable at low temperatures hydrocarbon phases (e.g. ethane) can be synthesized at high P-T conditions \cite{Kolesnikov2009, Lobanov2013}, that have been argued to stabilize at high temperature and via catalytic reaction \cite{Spanu2011}.

To address the structure and the composition of hydrocarbons at high pressures we have performed a combined experimental, using x-ray diffraction (XRD) and Raman spectroscopy, and theoretical, using the USPEX  ab-initio evolutionary structural search algorithm \cite{Zhu2012,Naumova2019}, study of both methane and ethane up to megabar pressures at room temperature (RT).  For ethane we have determined the crystallization point, for room temperature, at 2.7 GPa and also the low pressure crystal structure (Phase A), in agreement with Ref \onlinecite{Podsiadlo2017}.  This crystal structure is orientationally disordered (plastic phase) and deviates from the known crystal structures for ethane at low temperatures \cite{Klimenko2008,Nes1978}. In addition, a pressure induced phase transition has been identified, for the first time, at 13.6 GPa to a monoclinic phase B, the structure of which is solved based on a good agreement of the experimental results and theoretical predictions. This phase is isostructural with phase III, previously reported at low temperatures \cite{Klimenko2008}.  We have determined the equations of states (EOS) of ethane and methane up to Mbar pressures, which provides a solid basis for the discussion of their relative stability at high pressures.

\section{Methods}

\subsection{Experimental methods}
Methane and ethane (both with nominal purity better than 99.9995\%) were loaded into a diamond anvil cell (DAC) with the use of a  gas loading apparatus, were a gas pressure of  about  0.2 GPa was created. Small quantities of ruby and gold powder were also loaded, for determination of pressure through ruby luminescence  and gold EOS, respectively \cite{Syassen2008,Matsui2010}. XRD data were collected at the GeoSoilEnviroCARS (sector 13), APS, Chicago and Extreme Conditions Beamline P02.2 at DESY (Germany).   Cubic boron nitride (c-BN) gaskets were used for the Mbar XRD measurements. This allows larger sample volumes, given that both methane and ethane are week scatterers. Moreover, the use of c-BN gasket prevents \textquotedblleft contamination \textquotedblright  of XRD patterns from Rhenium peaks, which are usually present at high pressures \cite{Hirai2008,Sun2009}.    Raman studies were performed using 488 and 532 nm lines of a solid-state laser. Raman spectra were analyzed with a spectral resolution of 4 cm$^{-1}$ using a single-stage grating spectrograph equipped with a CCD array detector.

Integration of powder diffraction patterns to yield scattering intensity versus 2$\theta$ diagrams and initial analysis were performed using the DIOPTAS program \cite{Prescher2015}. Calculated XRD patterns were produced using the POWDER CELL program \cite{Kraus1996}, for the corresponding crystal structures according to the EOSs determined experimentally and theoretically in this study and assuming continuous Debye rings of uniform intensity. Rietveld refinements were performed using the GSAS software \cite{Larson2000}. Indexing of XRD patterns has been performed using the DICVOL program \cite{Boutlif2004} as implemented in the FullProf Suite.

\subsection{Computational  methods}
The computational search for stable phases of ethane under pressure was performed using the USPEX evolutionary algorithm \cite{Glass2006} in its version  \cite{Zhu2012} for molecular crystals. Force-field (UFF \cite{Rappe1992}) optimized geometry of ethane molecule was used as input for structure search. We explicitly considered structures with Z=2, 4, 8, 10, 12, 14, 16 molecules per unit cell,  in all possible space groups. Generation size was set to 80 structures, stopping criterion of 6 generations was used. External pressure was set to 30 GPa during the search.
Each structure in USPEX search was relaxed using VASP \cite{Kresse1993,Kresse1996,Kresse1996a} package with PBE \cite{perdew1996} density functional (with D3 dispersion correction \cite{Grimme2010}) and PAW pseudopotentials \cite{kresse1999} in several consecutive steps with increasing precision, planewave basis set cutoff and number of points on k-grid: final, most precise step was done with cutoff of 600 eV and 2$\pi$X0.09 \AA$^{-1}$ reciprocal space resolution (as defined in the original USPEX paper \cite{Glass2006}). Fitting of Birch-Murnaghan EOS was done by relaxing the predicted ethane structure at 10 equidistant pressure values in 10-100 GPa range with 600 eV cutoff and 0.5 \AA$^{-1}$ k-point spacing; obtained equilibrium cell volumes and energies were used for a three-parameter fit.

\section{Results}

\subsection{High pressure study of Ethane}
\subsubsection{Raman spectroscopy}

Figure 1 shows nonpolarized Raman spectra of ethane at selected pressures up to 40 GPa.  Frequency pressure plots are shown in Figure 2. Ethane crystallizes at about 2 GPa, determined by the appearance of lattice modes above solidification pressure. The Raman spectrum can be divided into 3 main spectral ranges: i) from 2850 to 3100 cm$^{-1}$ attributed to C-H stretching vibrations \cite{Helvoort1987}, ii) from 1000 to 1500cm$^{-1}$ attributed to  C-C stretching and C-H bending and iii) bellow 600 cm$^{-1}$ arising from lattice modes.

\begin{figure}[ht]
\centering
\includegraphics[width=\linewidth]{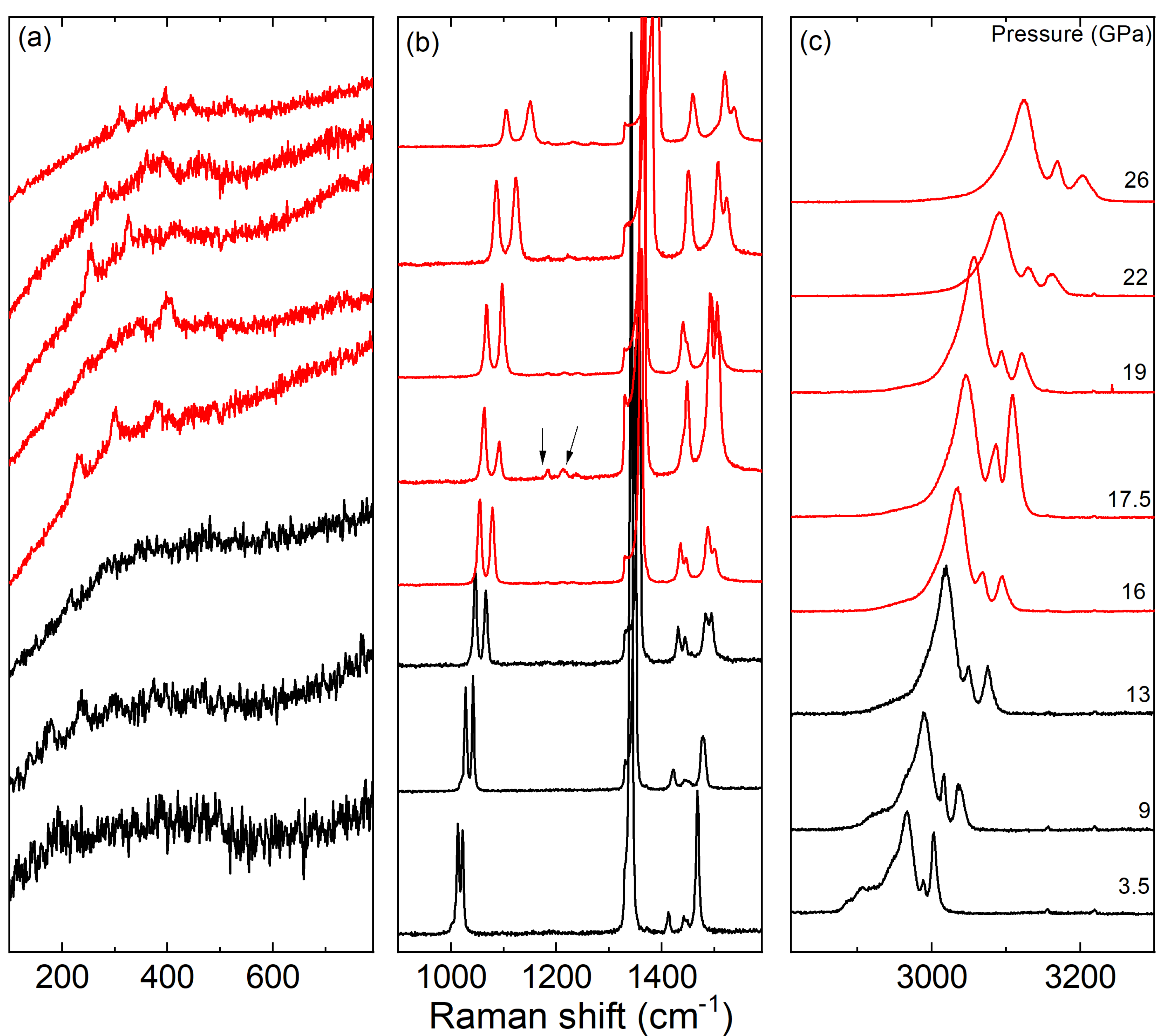}
\caption{Raman spectra of ethane at selected pressures: a) lattice modes, b) C-C stretching and C-H bending modes and c) C-H stretching modes. The Raman spectra of the A and B phases are shown by black and  red, respectively. }
\end{figure}

At above 13 GPa a phase transition has been observed manifested by the decrease of the number of high-frequency C-H stretching modes with the parallel appearance of   C-H bending modes at 1200 cm$^{-1}$ (indicate by arrows in Fig. 1). Moreover an increase of the number of the lattice modes (see Fig. 2) indicates a lower symmetry unit cell. No indication  of additional phase transitions was observed up to the highest pressure of this study (40 GPa).

\begin{figure}[ht]
\centering
\includegraphics[width=\linewidth]{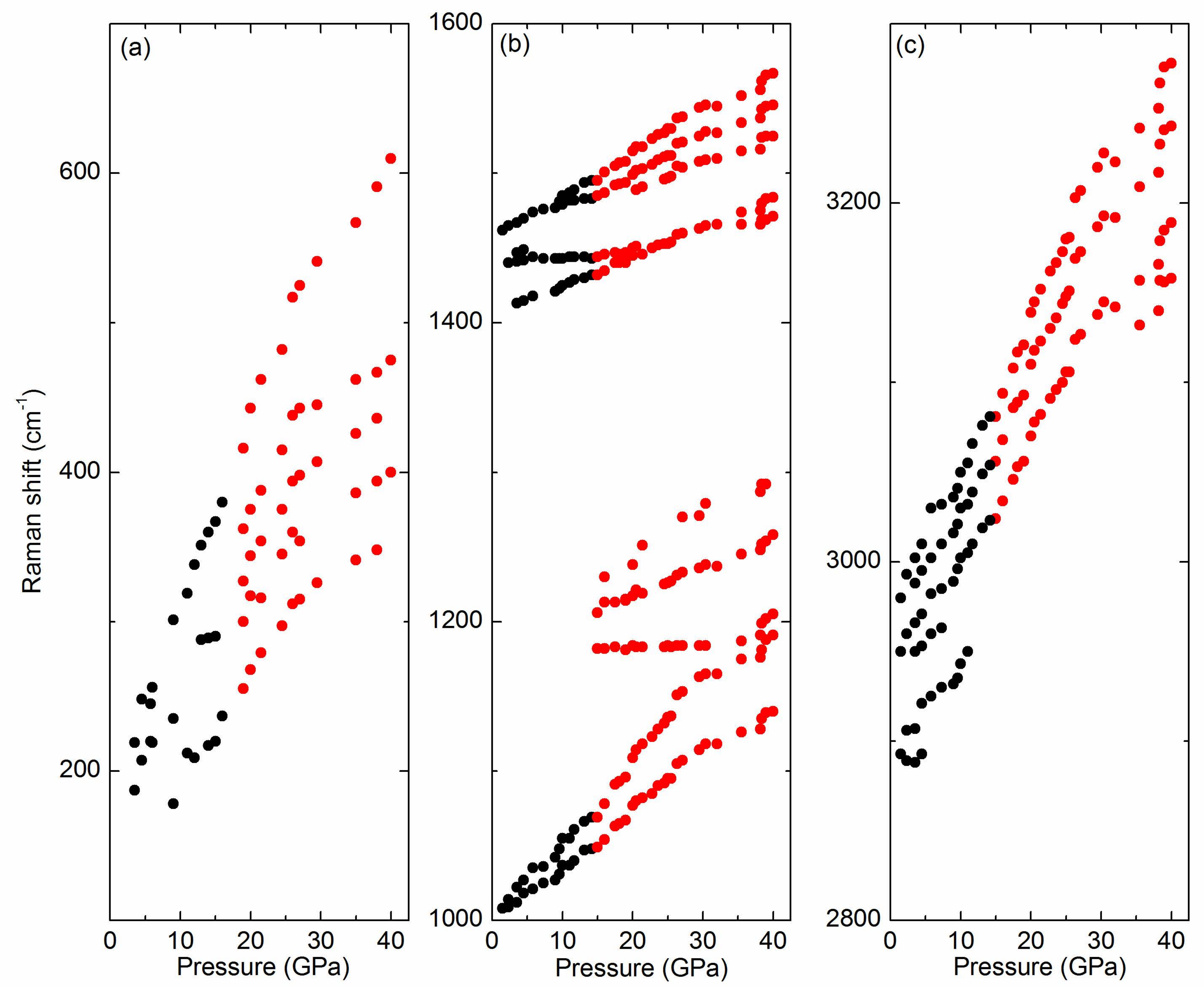}
\caption{ Frequency vs pressure plots of the ethane Raman modes: a) lattice modes, b) C-C stretching and C-H bending modes and c) C-H stretching modes. The Raman mode frequencies of  the A and B phases are shown by black and  red, respectively.}
\end{figure}

\subsubsection{X-ray diffraction}
Representative XRD patterns of ethane at various pressure are shown in figure 3. The crystalline phase (indicated as phase A in our study and phase IV in Ref. \onlinecite{Podsiadlo2017}) of ethane at the lower pressure range (from crystallization up to 13 GPa) cannot be indexed with the known crystalline phases at low temperatures. Indeed, both the plastic BCC (phase I) and the monoclinic ordered (phase III) phases reported previously \cite{Klimenko2008,Nes1978} have XRD patterns distinct from those observed in our study. With pressure increase a clear indication of a phase transition to a high pressure phase (indicated as phase B) has been observed at 13.4 GPa , in perfect agreement with Raman spectroscopy.  No other phase transition has been observed up to 110 GPa.

\begin{figure}[ht]
\centering
\includegraphics[width=\linewidth]{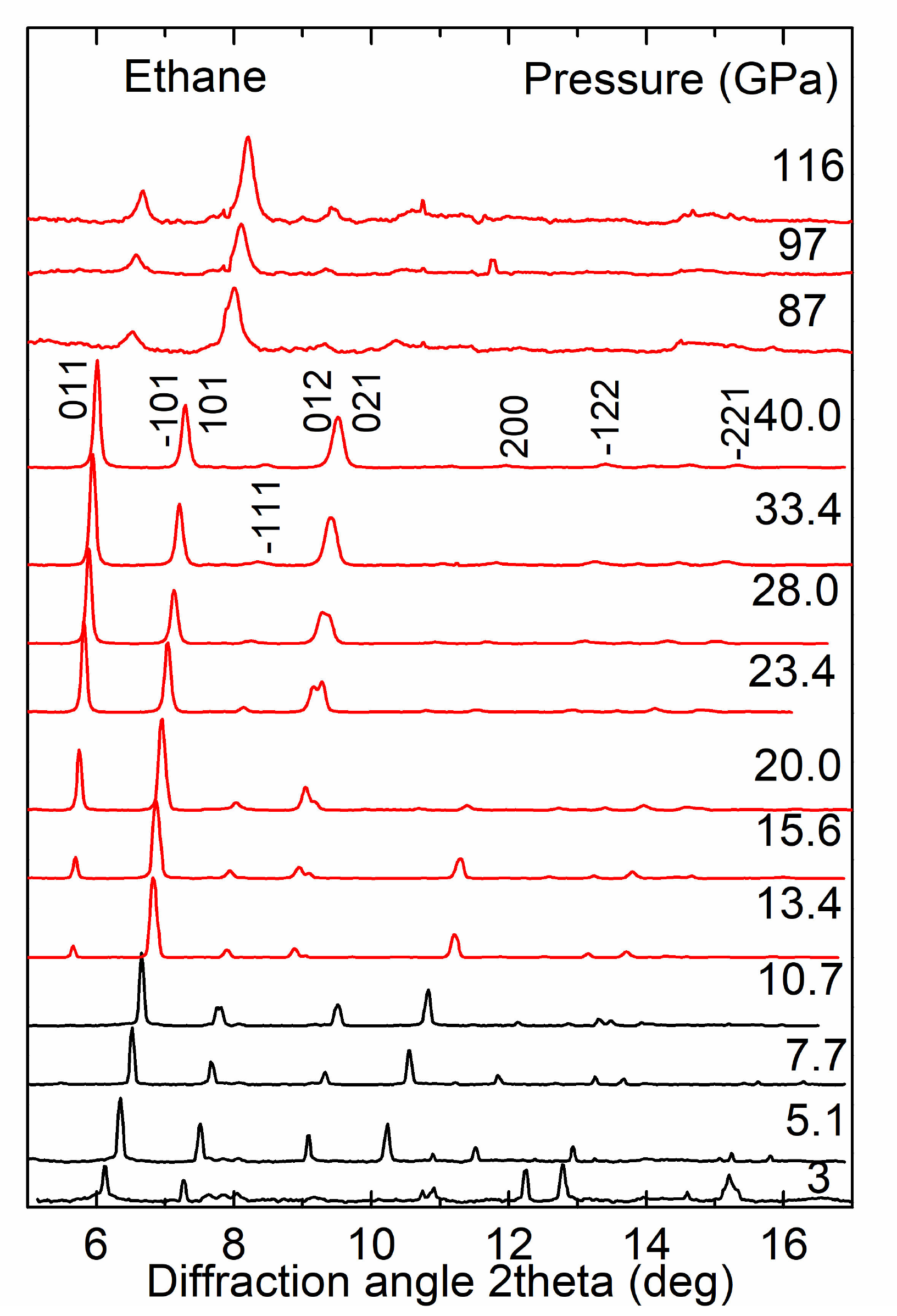}
\caption{ Selected XRD patterns of ethane at various pressures measured on pressure increase. The XRD patterns of  the A and B phases are shown by black and  red, respectively. The corresponding Miller indices for phase B are also noted. The X-ray wavelength is $\lambda$=0.3344 \AA. }
\end{figure}

In order to solve the crystal structures of both phases of ethane we have performed a combined ab-initio and experimental (indexing) procedure. We would like to point out that from experimental data only carbon positions can be extracted from experimental data due to the low scattering cross section of hydrogen atoms. The results of the  analysis of the experimental results are as follows: (a) Phase A can be  indexed with a tetragonal cell (a=b=5.11 c=3.74 \AA  {} at 5 GPa) in agreement with Ref. \onlinecite{Podsiadlo2017}  ,  (b) Phase B is determined as a monoclinic structure (space group (S.G.) $P2_1/n$ (14) Z=2) with   $a$=3.40 $b$=4.66 $c$=4.76 \AA {} $\beta$=90.45$^o$ at 20 GPa, and (c) A and B phases should be closely related, most probably holding an orientation order-disorder relation with higher symmetry carbon positions in phase A. This is further justified by Raman spectroscopy, where an increased number of the low frequency lattice modes has been observed after the phase transition. The lattice mode frequencies of phase B seem to be on the extension of phase A dependencies, see Fig. 2(a).  The increased number of the lattice modes and the appearance of the C-H bending modes suggest the symmetry lowering.

Computational search for  ethane above 14 GPa produced 486 enthalpy-ranked structures, which were analyzed for matching the experimental XRD and cell parameters in order of enthalpy increase. The most stable (lowest energy) structure found during the computational search immediately returned lattice parameters and XRD peak positions closely matching ones of experimentally found phase B. The next, theoretically predicted, low energy structure has an energy of 1.245 kcal/mol above this structure. Calculated lattice parameters of the lowest energy structure are $a$=3.22,$b$=4.49,$c$=4.52, $\beta$=90.47$^o$ with $P2_1/n$ S.G. and Z=2 at 30 GPa. We consider the similarity with the experimentally determined lattice parameters and S.G. sufficient to uniquely identify this structure as the phase B. Interestingly, this structure was found to be the stable one in several independent searches.  The close agreement between the experimentally and computationally determined structures of phase B of ethane give us confidence that this should be the correct structure of ethane above 13 GPa. This is further justified by the fair agreement between the experimentally and computationally determined pressure dependance of the  lattice parameters and the EOSs for phase B, see Figs. 4 (a) and (b).

Our XRD experiments and theoretical calculations suggest that phase B is orientantionally ordered BCT-like structure (see insets in Fig. 4) in which the ethane molecule at the center of the unit cell has an orientation (the direction of the C-C bond) perpendicular to the orientation of the molecules at the corners.  This crystal structure resembles very much the monoclinic Phase III known from LT studies \cite{Klimenko2008}. At 70 K and ambient pressure Phase III has lattice constants a=4.226; b=5.623; c=5.845 \AA; {}$\beta$=90.41$^o$ with the same SG ($P2_1/n$) of Phase B. Moreover, arrangement of ethane molecules is the same in both structures.  Thus, it is plausible to assume, although no experimental data are available in the intermediate temperature and pressure region, that phase B is identical with the LT phase III.

\begin{figure}[ht]
\centering
\includegraphics[width=\linewidth]{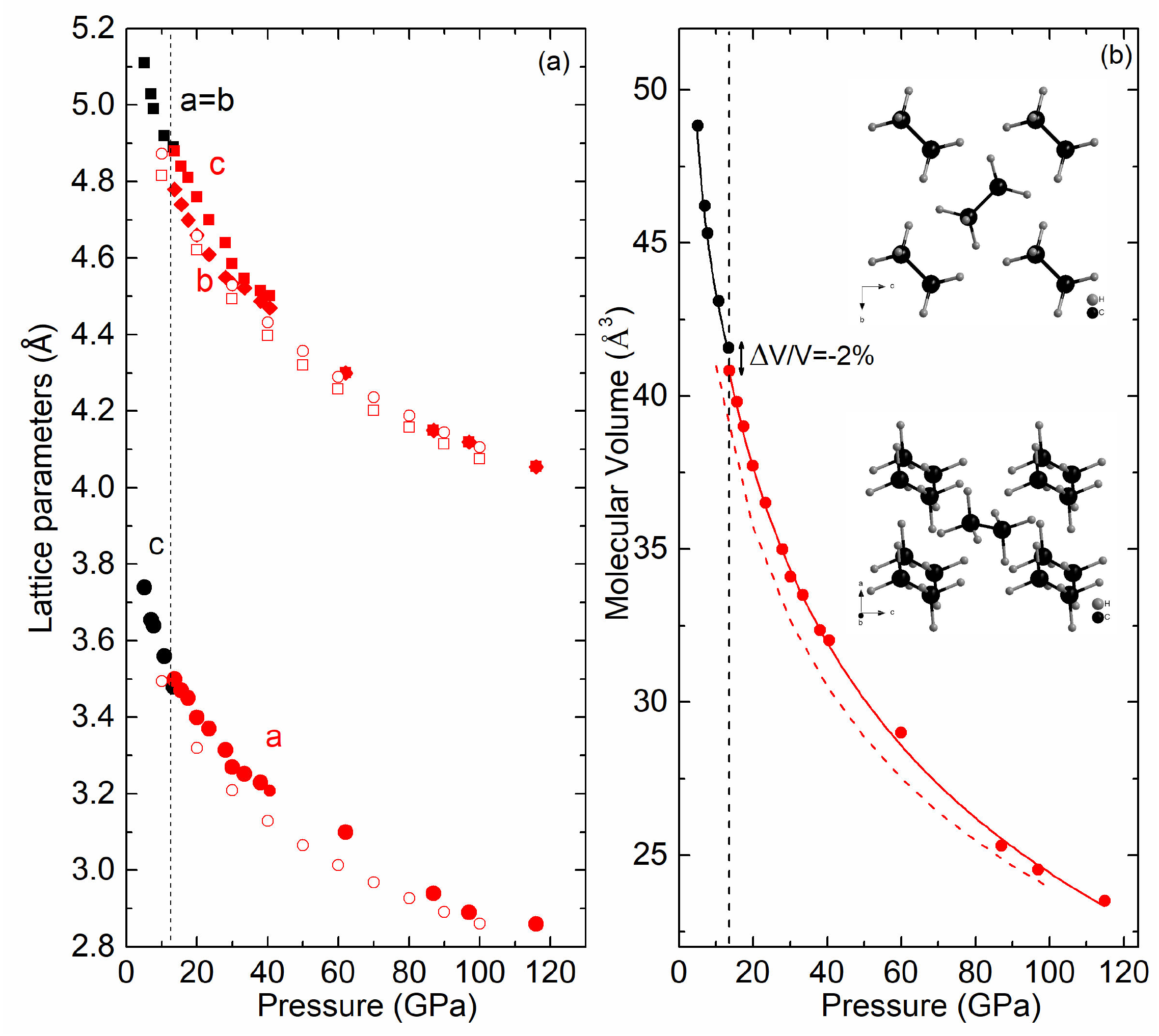}
\caption{Pressure dependence of: (a) lattice parameters  and (b) molecular volume of ethane.  The dashed vertical lines denote the critical pressure for the phase transition. In (a) experimental and calculated results are plotted with solid and open symbols, respectively. The solid lines in (b) are unweighted third-order Birch-Murnaghan EOS fit to the experimental data points \cite{Birch1978a}. The calculated EOS of phase B is shown with dashed red line in (b). The insets in (b) are schematic representations of the B phase of ethane. }
\end{figure}

Using the above-mentioned structures, we have obtained the lattice parameters and the  EOS of ethane up to 110 GPa. The results are shown in figure 4. The pressure induced A$\rightarrow$B phase transition is accompanied by a slight ($\approx$-2\%) volume drop.  As it can be clearly seen from Fig. 4, the lattice parameters b and c become practically identical above 40 GPa while the $\beta$ monoclinic angle remains very close to 90$^o$, thus making phase B effectively tetragonal under pressure. The continuity of the lattice parameters (c$\rightarrow$a and a,b$\rightarrow$b,c) further justifies the argument of the close relation between the two phases. We conducted unweighted fits to the experimental and calculated P-V data using a third-order Birch-Murnaghan EOS \cite{Birch1978a} and determined the bulk modulus $B$ and its first derivative $B'$ at  the experimental onset pressure for the A and B phases. The lattice, structural, and EOS parameters  obtained in this way are given in Table I.

\begin{table*}[tb]\centering
\scriptsize
\caption{Experimental and calculated structural parameters of A and B ethane  at selected pressures: space group (SG), number of formula units in the unit cell Z, lattice parameters,  molecular volume, bulk modulus B and its pressure derivative B' at  the experimental onset pressure, Wyckoff site and the corresponding coordinates as determined by ab-initio calculations.}
\medskip
\setlength{\extrarowheight}{1pt}
\begin{ruledtabular}
\begin{tabular}{cccccccccccccc}
P(GPa)& SG& Z & $a$({\AA}) & $b$({\AA}) & $c$({\AA}) &$\beta$ (deg)&  V(\AA$^3$)& B(GPa)&  B' & WP & x&y&z \\\hline
7 (exp)& $P4_2/mnm$& 2& 5.028(1)&5.028(1)&3.654(1))&&46.25(5)& 25.9(8)&9.2 (7) & &   &    &\\
 5.9 (Ref. 19)& $P4_2/mnm$&& 5.092(2) &5.092(2)    &3.675(18)&&  47.65   &      &&& \\\hline
20 (exp.) &$P2_1/n$&2&3.40(2)&	4.661(5)&	4.762(4)&90.45&37.71(11)&71.1(9)& 3.5 (17)& C(4e)&0.0586& 0.3890& 0.8897   \\
30 (cal.)& $P2_1/n$&2&3.22   & 4.49      & 4.52&     90.47&39.1     &64     & 4.1     &H(4e) &0.1135 &0.8673& 0.6282 \\
&      &  &     &     &     &       & &&&H(4e) &0.5630& 0.6754& 0.55530 \\
&      &  &     &     &     &       & &&&H(4e) &0.4443& 0.0558& 0.1745  \\
\end{tabular}
\end{ruledtabular}
\end{table*}

Figure 5 shows the proposed phase diagram of ethane summarizing the findings of this work. Phase B, representing an orientationally ordered molecular structure is stable in the whole explored pressure range up to 116 GPa in agreement with the theoretical predictions \cite{Naumova2019}.

\begin{figure}[ht]
\centering
\includegraphics[width=\linewidth]{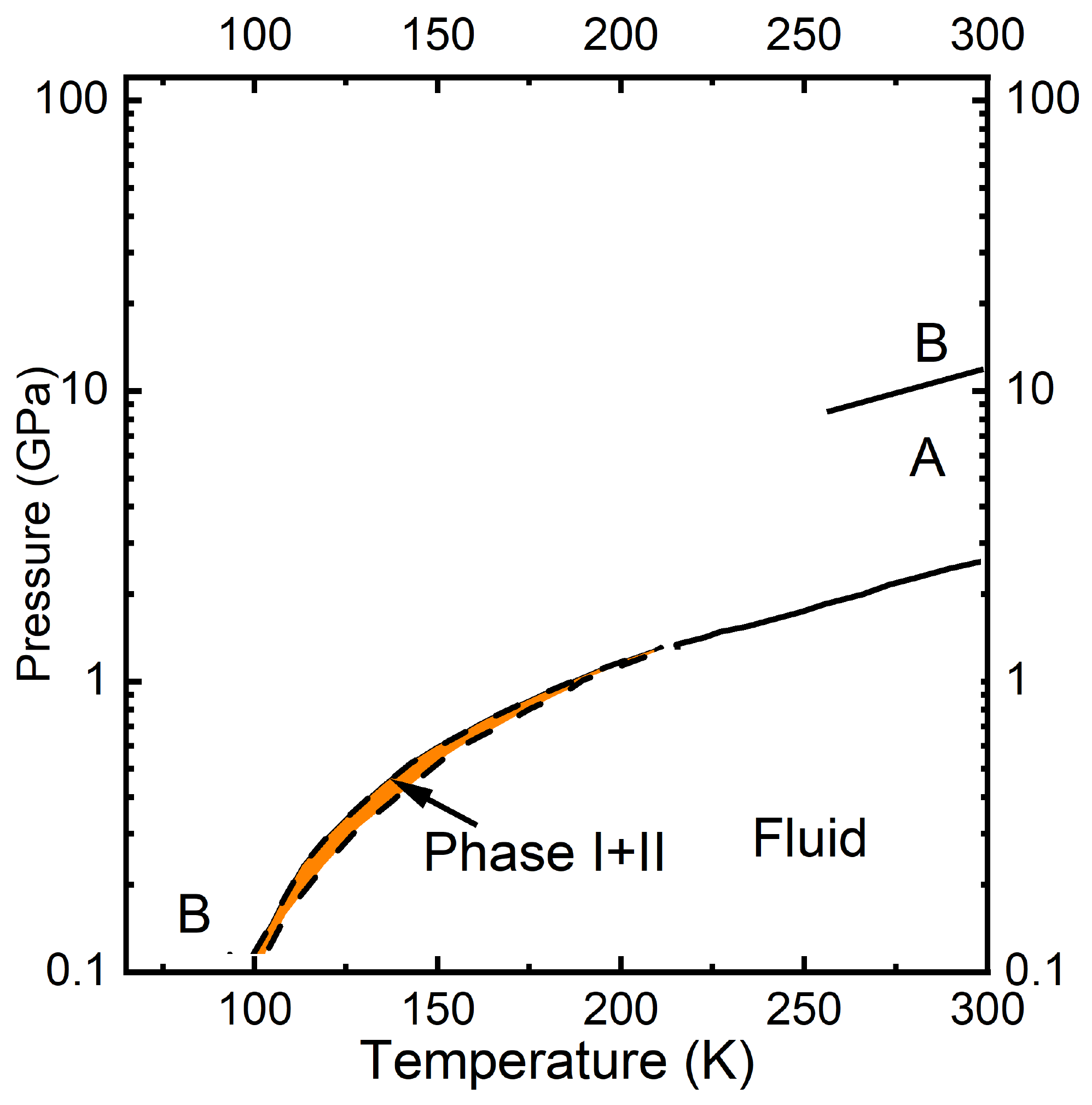}
\caption{Proposed P-T phase diagram of ethane according to the results of this study and Ref. \onlinecite{Podsiadlo2017}. }
\end{figure}

\subsection{High pressure study of methane}
\subsubsection{X-ray diffraction}
The XRD data of methane up to 117 GPa collected in this work are consistent with the I-A-B sequence of phase transformations. Concomitant Raman spectra measurements indicated that methane was in phase pre-B above 12 GPa. Our XRD data (Fig. 6) show that in the whole investigated pressure range up to 120 GPa the data can be indexed within a cubic structure.

\begin{figure}[ht]
\centering
\includegraphics[width=\linewidth]{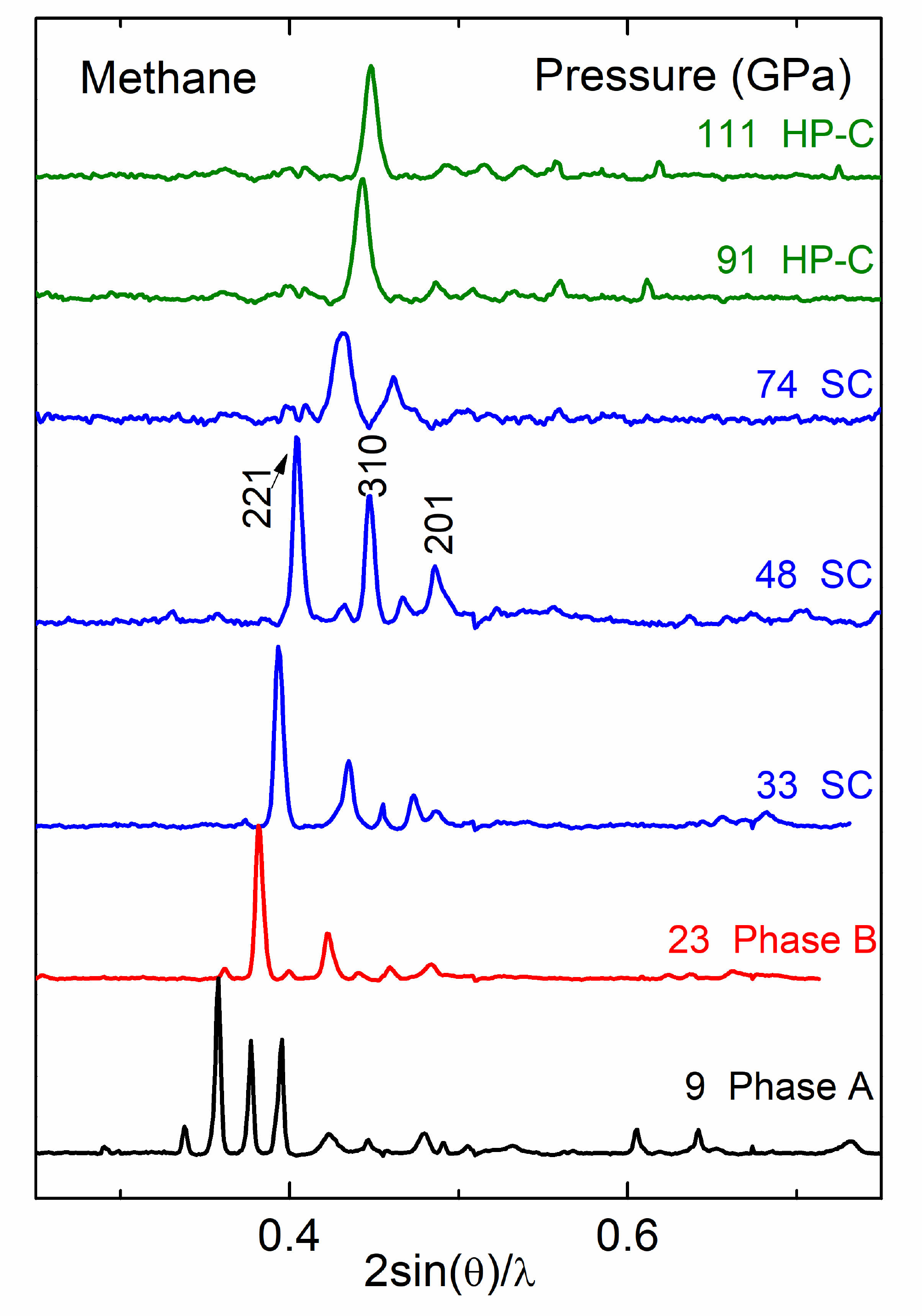}
\caption{Selected XRD patterns of methane at various pressures measured on pressure increase. The corresponding Miller indices for SC phase of methane are also noted.}
\end{figure}

These observations suggest that the simple cubic structure determined by the carbon sublattice remains stable under compression. On the other hand, previous Raman investigations show that there may be a variety of ordered phases which likely occur due to different modes of orientational ordering \cite{Chen2011,Bini1995,Bini1997}, which depend both on the sample history and pressure. For these structures the H-positions are unknown while C positions were determined in Ref. 9. for phase B.  The investigated here cubic pre-B methane, which is stable to very high pressures (see also Refs. \onlinecite{Hirai2008,Sun2009}), is a good proxy to determine the EOS at these conditions. We argue that pre-B, B and  HP phases of methane are close in energy because they show metastablity.

\subsubsection{Raman spectroscopy}

Our Raman data measured concomitantly with XRD agree well with previous observations in phases I, A, B, and pre-B \cite{Hirai2008,Proctor2017}. We detected a simple Raman spectrum of the C-H stretch modes, which consists of two major fundamentals ($\nu_1$ and $\nu_3$) up to the highest pressure of investigation, in agreement with Ref. 14.

\section{Discussion}
Methane and ethane both are the simplest and hence the most fundamental hydrocarbons. Thus, it is vital to understand their relative thermodynamic stability at different compressions. As already mentioned in the introduction, ethane is one of the most predictable products of chemical reactivity of methane at extreme pressures and temperatures. In order to address the issue of the relative stability of methane vs ethane, and  given that the PV enthalpy term plays the most important role at high pressure, we compare the volume per molecule of  methane and ethane. This is done using  the simple reaction 2CH$_4$$\Rightarrow$C$_2$H$_6$ + H$_2$. Figure 7 shows the combined EOS of : (a) doubled molecular volume (Vpm) of methane (b) sum of the molecular volume of ethane Vpm plus hydrogen Vpm according to Loubeyre $et al.$ \cite{Loubeyre1996}  As it can be clearly seen,  ethane and methane show similar compressibility, which results to similar EOSs. In details,  ethane seems more stable up to 80 GPa, while both have similar stabilities above this pressure where entropy term is likely less important. Two volumes are similar, within the experimental error, above 80 GPa and  up to 120 GPa $i.e.$ 2Vpm(CH$_4$) and Vpm(C$_2$H$_6$+H$_2$) are almost the same.

\begin{figure}[ht]
\centering
\includegraphics[width=\linewidth]{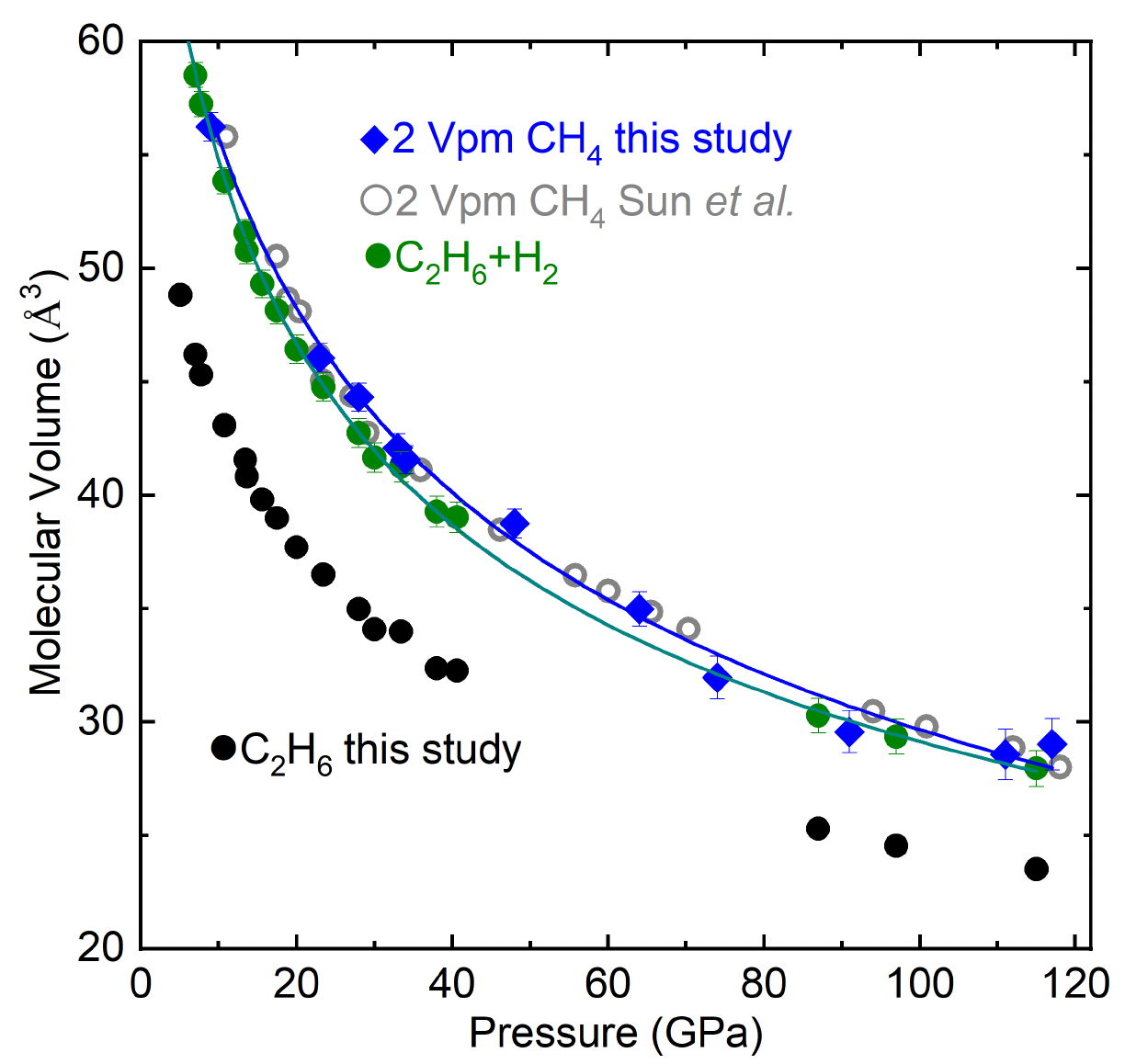}
\caption{Comparative equations of state of:  ethane (black symbols), and methane (blue symbols our data and red symbols  from Ref. \onlinecite{Sun2009}). The EOS of methane is compared to a combined EOS of ethane (this work) and molecular hydrogen \cite{Loubeyre1996}, which has the same relative composition of carbon and hydrogen.}
\end{figure}

High-temperature (HT) studies unambiguously reveal the formation of ethane, originated from methane, under high pressure-temperature conditions \cite{Kolesnikov2009, Lobanov2013}. In this case, HT commonly needed to overcome energy barriers arising from bond breaking. Moreover, previous theoretical studies strongly suggest dissociation of methane towards ethane at 95 GPa \cite{Gao2010}/ From our study, the PV terms look similar for methane and ethane at high pressures, or at least the difference between PV terms remains stable. Since it is plausible to assume that internal energy terms are also similar, then this implies that the relative enthalpy $\Delta$H difference between methane and ethane remains stable. Consequently, the transformation of methane to ethane cannot be justified by simple enthalpy arguments.

In order to overcome this discrepancy we turn our attention to Gibbs free energy   G(p,T) = H-TS. Since for a favorable transformation  a negative $\Delta$G is needed, then the entropy of hydrogen SH$_2$ should have the key importance, if we accept that $\Delta$G=-$TS$H$_2$ , $i.e.$ $S_{meth}$=$S_{ethan}$, especially, at room temperature. Indeed an increasing $S$ of hydrogen with pressure would explain the transformation at high pressures and temperatures given that kinetics restrictions can be overcome by HT. The situation resembles very much the diamond to graphite conversion. At   RT diamond has a negligibly higher $H$ than graphite but also a reasonably higher $G$ due to lower $S$, therefore the conversion C(d)$\Rightarrow$C(g)  is energetically favorable although kinetics prevents this transformation at RT. High-temperatures are needed in order to overcome the barriers.

\section{Summary}
High-pressure phase transitions of ethane and methane  have been investigated by a combined experimental (using X-ray diffraction and Raman spectroscopy) and computational (USPEX) study up to above 100 GPa. In the case of ethane, a pressure induced phase transition, was observed above  13.6 GPa, supported by both XRD and Raman measurements. The structure of this high-pressure phase, namely phase B, was definitively determined using a combined excremental and theoretical approach; phase B appears to be isostructural to the low-temperature phase III at ambient pressure known previously.   This phase remains stable up to the highest pressure of this study and a P-T phase diagram of ethane  is proposed. In the case of methane, our XRD measurements revealed that in our experiments methane followed the phase sequence I$\rightarrow$A$\rightarrow$pre-B and pre-B seems to be stable to the highest pressure reached.  Comparison of the ethane and methane EOSs, after adjusting the volume of ethane with hydrogen volume, revealed that there is no difference, outside the experimental error, between the two volumes $i.e.$ 2Vpm(CH$_4$) and Vpm(C$_2$H$_6$+H$_2$) are almost the same. We conclude that an increasing $S$ of hydrogen with pressure is the plausible explanation for the observed transformation of  methane to ethane under  high pressure-temperature conditions.

\acknowledgments
A.F.G. and E.S. acknowledge support from the Army Research Office (Grants No. 56122-CH-H and No. 71650-CH) and the Deep Carbon Observatory. S.S.L. acknowledges the support of the Helmholtz Young Investigators Group CLEAR (VH‐NG‐1325).  Part of this work was performed at GeoSoilEnviroCARS (The University of Chicago, Sector 13), Advanced Photon Source (APS), Argonne National Laboratory. GeoSoilEnviroCARS is supported by the National Science Foundation - Earth Sciences (EAR-1634415) and Department of Energy-GeoSciences (DE-FG02-94ER14466). This research used resources of the Advanced Photon Source, a U.S. Department of Energy (DOE) Office of Science User Facility operated for the DOE Office of Science by Argonne National Laboratory under Contract No. DE-AC02-06CH11357. We acknowledge DESY (Hamburg, Germany), a member of the Helmholtz Association HGF, for the provision of experimental facilities. Part of this research was carried out at PETRA III (beamline P02.2). The research leading to these results has received funding from the European Community’s Seventh Framework Programme (FP7/2007-2013) under grant agreement n 312284.


\begin{thebibliography}{42}%
\makeatletter
\providecommand \@ifxundefined [1]{%
 \@ifx{#1\undefined}
}%
\providecommand \@ifnum [1]{%
 \ifnum #1\expandafter \@firstoftwo
 \else \expandafter \@secondoftwo
 \fi
}%
\providecommand \@ifx [1]{%
 \ifx #1\expandafter \@firstoftwo
 \else \expandafter \@secondoftwo
 \fi
}%
\providecommand \natexlab [1]{#1}%
\providecommand \enquote  [1]{``#1''}%
\providecommand \bibnamefont  [1]{#1}%
\providecommand \bibfnamefont [1]{#1}%
\providecommand \citenamefont [1]{#1}%
\providecommand \href@noop [0]{\@secondoftwo}%
\providecommand \href [0]{\begingroup \@sanitize@url \@href}%
\providecommand \@href[1]{\@@startlink{#1}\@@href}%
\providecommand \@@href[1]{\endgroup#1\@@endlink}%
\providecommand \@sanitize@url [0]{\catcode `\\12\catcode `\$12\catcode
  `\&12\catcode `\#12\catcode `\^12\catcode `\_12\catcode `\%12\relax}%
\providecommand \@@startlink[1]{}%
\providecommand \@@endlink[0]{}%
\providecommand \url  [0]{\begingroup\@sanitize@url \@url }%
\providecommand \@url [1]{\endgroup\@href {#1}{\urlprefix }}%
\providecommand \urlprefix  [0]{URL }%
\providecommand \Eprint [0]{\href }%
\providecommand \doibase [0]{http://dx.doi.org/}%
\providecommand \selectlanguage [0]{\@gobble}%
\providecommand \bibinfo  [0]{\@secondoftwo}%
\providecommand \bibfield  [0]{\@secondoftwo}%
\providecommand \translation [1]{[#1]}%
\providecommand \BibitemOpen [0]{}%
\providecommand \bibitemStop [0]{}%
\providecommand \bibitemNoStop [0]{.\EOS\space}%
\providecommand \EOS [0]{\spacefactor3000\relax}%
\providecommand \BibitemShut  [1]{\csname bibitem#1\endcsname}%
\let\auto@bib@innerbib\@empty
%</preamble>
\bibitem [{\citenamefont {Hubbard}(1981)}]{Hubbard1981}%
  \BibitemOpen
  \bibfield  {author} {\bibinfo {author} {\bibfnamefont {W.~B.}\ \bibnamefont
  {Hubbard}},\ }\href {\doibase 10.1126/science.214.4517.145} {\bibfield
  {journal} {\bibinfo  {journal} {Science}\ }\textbf {\bibinfo {volume}
  {214}},\ \bibinfo {pages} {145} (\bibinfo {year} {1981})}\BibitemShut
  {NoStop}%
\bibitem [{\citenamefont {Podolak}, \citenamefont {Weizman},\ and\
  \citenamefont {Marley}(1995)}]{Podolak1995}%
  \BibitemOpen
  \bibfield  {author} {\bibinfo {author} {\bibfnamefont {M.}~\bibnamefont
  {Podolak}}, \bibinfo {author} {\bibfnamefont {A.}~\bibnamefont {Weizman}}, \
  and\ \bibinfo {author} {\bibfnamefont {M.}~\bibnamefont {Marley}},\ }\href
  {\doibase https://doi.org/10.1016/0032-0633(95)00061-5} {\bibfield  {journal}
  {\bibinfo  {journal} {Planet. Space Sci.}\ }\textbf {\bibinfo {volume}
  {43}},\ \bibinfo {pages} {1517 } (\bibinfo {year} {1995})}\BibitemShut
  {NoStop}%
\bibitem [{\citenamefont {Ancilotto}\ \emph {et~al.}(1997)\citenamefont
  {Ancilotto}, \citenamefont {Chiarotti}, \citenamefont {Scandolo},\ and\
  \citenamefont {Tosatti}}]{Ancilotto1997}%
  \BibitemOpen
  \bibfield  {author} {\bibinfo {author} {\bibfnamefont {F.}~\bibnamefont
  {Ancilotto}}, \bibinfo {author} {\bibfnamefont {G.~L.}\ \bibnamefont
  {Chiarotti}}, \bibinfo {author} {\bibfnamefont {S.}~\bibnamefont {Scandolo}},
  \ and\ \bibinfo {author} {\bibfnamefont {E.}~\bibnamefont {Tosatti}},\ }\href
  {\doibase 10.1126/science.275.5304.1288} {\bibfield  {journal} {\bibinfo
  {journal} {Science}\ }\textbf {\bibinfo {volume} {275}},\ \bibinfo {pages}
  {1288} (\bibinfo {year} {1997})}\BibitemShut {NoStop}%
\bibitem [{\citenamefont {Kolesnikov}, \citenamefont {Kutcherov},\ and\
  \citenamefont {Goncharov}(2009)}]{Kolesnikov2009}%
  \BibitemOpen
  \bibfield  {author} {\bibinfo {author} {\bibfnamefont {A.}~\bibnamefont
  {Kolesnikov}}, \bibinfo {author} {\bibfnamefont {V.~G.}\ \bibnamefont
  {Kutcherov}}, \ and\ \bibinfo {author} {\bibfnamefont {A.~F.}\ \bibnamefont
  {Goncharov}},\ }\href {https://doi.org/10.1038/ngeo591} {\bibfield  {journal}
  {\bibinfo  {journal} {Nat. Geosci.}\ }\textbf {\bibinfo {volume} {2}},\
  \bibinfo {pages} {566} (\bibinfo {year} {2009})}\BibitemShut {NoStop}%
\bibitem [{\citenamefont {Gao}\ \emph {et~al.}(2010)\citenamefont {Gao},
  \citenamefont {Oganov}, \citenamefont {Ma}, \citenamefont {Wang},
  \citenamefont {Li}, \citenamefont {Li}, \citenamefont {Iitaka},\ and\
  \citenamefont {Zou}}]{Gao2010}%
  \BibitemOpen
  \bibfield  {author} {\bibinfo {author} {\bibfnamefont {G.}~\bibnamefont
  {Gao}}, \bibinfo {author} {\bibfnamefont {A.~R.}\ \bibnamefont {Oganov}},
  \bibinfo {author} {\bibfnamefont {Y.}~\bibnamefont {Ma}}, \bibinfo {author}
  {\bibfnamefont {H.}~\bibnamefont {Wang}}, \bibinfo {author} {\bibfnamefont
  {P.}~\bibnamefont {Li}}, \bibinfo {author} {\bibfnamefont {Y.}~\bibnamefont
  {Li}}, \bibinfo {author} {\bibfnamefont {T.}~\bibnamefont {Iitaka}}, \ and\
  \bibinfo {author} {\bibfnamefont {G.}~\bibnamefont {Zou}},\ }\href {\doibase
  10.1063/1.3488102} {\bibfield  {journal} {\bibinfo  {journal} {J. Chem.
  Phy.}\ }\textbf {\bibinfo {volume} {133}},\ \bibinfo {pages} {144508}
  (\bibinfo {year} {2010})}\BibitemShut {NoStop}%
\bibitem [{\citenamefont {Lobanov}\ \emph {et~al.}(2013)\citenamefont
  {Lobanov}, \citenamefont {Chen}, \citenamefont {Chen}, \citenamefont {Zha},
  \citenamefont {Litasov}, \citenamefont {Mao},\ and\ \citenamefont
  {Goncharov}}]{Lobanov2013}%
  \BibitemOpen
  \bibfield  {author} {\bibinfo {author} {\bibfnamefont {S.~S.}\ \bibnamefont
  {Lobanov}}, \bibinfo {author} {\bibfnamefont {P.-N.}\ \bibnamefont {Chen}},
  \bibinfo {author} {\bibfnamefont {X.-J.}\ \bibnamefont {Chen}}, \bibinfo
  {author} {\bibfnamefont {C.-S.}\ \bibnamefont {Zha}}, \bibinfo {author}
  {\bibfnamefont {K.~D.}\ \bibnamefont {Litasov}}, \bibinfo {author}
  {\bibfnamefont {H.-K.}\ \bibnamefont {Mao}}, \ and\ \bibinfo {author}
  {\bibfnamefont {A.~F.}\ \bibnamefont {Goncharov}},\ }\href
  {https://doi.org/10.1038/ncomms3446} {\bibfield  {journal} {\bibinfo
  {journal} {Nat. Commun.}\ }\textbf {\bibinfo {volume} {4}},\ \bibinfo {pages}
  {2446} (\bibinfo {year} {2013})}\BibitemShut {NoStop}%
\bibitem [{\citenamefont {Moses}, \citenamefont {Rages},\ and\ \citenamefont
  {Pollack}(1995)}]{Moses1995}%
  \BibitemOpen
  \bibfield  {author} {\bibinfo {author} {\bibfnamefont {J.~I.}\ \bibnamefont
  {Moses}}, \bibinfo {author} {\bibfnamefont {K.}~\bibnamefont {Rages}}, \ and\
  \bibinfo {author} {\bibfnamefont {J.~B.}\ \bibnamefont {Pollack}},\ }\href
  {\doibase https://doi.org/10.1006/icar.1995.1022} {\bibfield  {journal}
  {\bibinfo  {journal} {Icarus}\ }\textbf {\bibinfo {volume} {113}},\ \bibinfo
  {pages} {232 } (\bibinfo {year} {1995})}\BibitemShut {NoStop}%
\bibitem [{\citenamefont {Hirai}\ \emph {et~al.}(2008)\citenamefont {Hirai},
  \citenamefont {Konagai}, \citenamefont {Kawamura}, \citenamefont {Yamamoto},\
  and\ \citenamefont {Yagi}}]{Hirai2008}%
  \BibitemOpen
  \bibfield  {author} {\bibinfo {author} {\bibfnamefont {H.}~\bibnamefont
  {Hirai}}, \bibinfo {author} {\bibfnamefont {K.}~\bibnamefont {Konagai}},
  \bibinfo {author} {\bibfnamefont {T.}~\bibnamefont {Kawamura}}, \bibinfo
  {author} {\bibfnamefont {Y.}~\bibnamefont {Yamamoto}}, \ and\ \bibinfo
  {author} {\bibfnamefont {T.}~\bibnamefont {Yagi}},\ }\href {\doibase
  10.1088/1742-6596/121/10/102001} {\bibfield  {journal} {\bibinfo  {journal}
  {J. Phys. Conf. Ser.}\ }\textbf {\bibinfo {volume} {121}},\ \bibinfo {pages}
  {102001} (\bibinfo {year} {2008})}\BibitemShut {NoStop}%
\bibitem [{\citenamefont {Maynard-Casely}\ \emph {et~al.}(2010)\citenamefont
  {Maynard-Casely}, \citenamefont {Bull}, \citenamefont {Guthrie},
  \citenamefont {Loa}, \citenamefont {McMahon}, \citenamefont {Gregoryanz},
  \citenamefont {Nelmes},\ and\ \citenamefont {Loveday}}]{Casely2010}%
  \BibitemOpen
  \bibfield  {author} {\bibinfo {author} {\bibfnamefont {H.~E.}\ \bibnamefont
  {Maynard-Casely}}, \bibinfo {author} {\bibfnamefont {C.~L.}\ \bibnamefont
  {Bull}}, \bibinfo {author} {\bibfnamefont {M.}~\bibnamefont {Guthrie}},
  \bibinfo {author} {\bibfnamefont {I.}~\bibnamefont {Loa}}, \bibinfo {author}
  {\bibfnamefont {M.~I.}\ \bibnamefont {McMahon}}, \bibinfo {author}
  {\bibfnamefont {E.}~\bibnamefont {Gregoryanz}}, \bibinfo {author}
  {\bibfnamefont {R.~J.}\ \bibnamefont {Nelmes}}, \ and\ \bibinfo {author}
  {\bibfnamefont {J.~S.}\ \bibnamefont {Loveday}},\ }\href {\doibase
  10.1063/1.3455889} {\bibfield  {journal} {\bibinfo  {journal} {J. Chem.
  Phy.}\ }\textbf {\bibinfo {volume} {133}},\ \bibinfo {pages} {064504}
  (\bibinfo {year} {2010})}\BibitemShut {NoStop}%
\bibitem [{\citenamefont {Maynard-Casely}\ \emph {et~al.}(2014)\citenamefont
  {Maynard-Casely}, \citenamefont {Lundegaard}, \citenamefont {Loa},
  \citenamefont {McMahon}, \citenamefont {Gregoryanz}, \citenamefont {Nelmes},\
  and\ \citenamefont {Loveday}}]{Casely2014}%
  \BibitemOpen
  \bibfield  {author} {\bibinfo {author} {\bibfnamefont {H.~E.}\ \bibnamefont
  {Maynard-Casely}}, \bibinfo {author} {\bibfnamefont {L.~F.}\ \bibnamefont
  {Lundegaard}}, \bibinfo {author} {\bibfnamefont {I.}~\bibnamefont {Loa}},
  \bibinfo {author} {\bibfnamefont {M.~I.}\ \bibnamefont {McMahon}}, \bibinfo
  {author} {\bibfnamefont {E.}~\bibnamefont {Gregoryanz}}, \bibinfo {author}
  {\bibfnamefont {R.~J.}\ \bibnamefont {Nelmes}}, \ and\ \bibinfo {author}
  {\bibfnamefont {J.~S.}\ \bibnamefont {Loveday}},\ }\href {\doibase
  10.1063/1.4903813} {\bibfield  {journal} {\bibinfo  {journal} {J. Chem.
  Phy.}\ }\textbf {\bibinfo {volume} {141}},\ \bibinfo {pages} {234313}
  (\bibinfo {year} {2014})}\BibitemShut {NoStop}%
\bibitem [{\citenamefont {Umemoto}\ \emph {et~al.}(2002)\citenamefont
  {Umemoto}, \citenamefont {Yoshii}, \citenamefont {Akahama},\ and\
  \citenamefont {Kawamura}}]{Umemoto2002}%
  \BibitemOpen
  \bibfield  {author} {\bibinfo {author} {\bibfnamefont {S.}~\bibnamefont
  {Umemoto}}, \bibinfo {author} {\bibfnamefont {T.}~\bibnamefont {Yoshii}},
  \bibinfo {author} {\bibfnamefont {Y.}~\bibnamefont {Akahama}}, \ and\
  \bibinfo {author} {\bibfnamefont {H.}~\bibnamefont {Kawamura}},\ }\href
  {\doibase 10.1088/0953-8984/14/44/355} {\bibfield  {journal} {\bibinfo
  {journal} {J. Phys.: Condens. Matter}\ }\textbf {\bibinfo {volume} {14}},\
  \bibinfo {pages} {10675} (\bibinfo {year} {2002})}\BibitemShut {NoStop}%
\bibitem [{\citenamefont {Sun}\ \emph {et~al.}(2009)\citenamefont {Sun},
  \citenamefont {Yi}, \citenamefont {Wang}, \citenamefont {Shu}, \citenamefont
  {Sinogeikin}, \citenamefont {Meng}, \citenamefont {Shen}, \citenamefont
  {Bai}, \citenamefont {Li}, \citenamefont {Liu}, \citenamefont {kwang Mao},\
  and\ \citenamefont {Mao}}]{Sun2009}%
  \BibitemOpen
  \bibfield  {author} {\bibinfo {author} {\bibfnamefont {L.}~\bibnamefont
  {Sun}}, \bibinfo {author} {\bibfnamefont {W.}~\bibnamefont {Yi}}, \bibinfo
  {author} {\bibfnamefont {L.}~\bibnamefont {Wang}}, \bibinfo {author}
  {\bibfnamefont {J.}~\bibnamefont {Shu}}, \bibinfo {author} {\bibfnamefont
  {S.}~\bibnamefont {Sinogeikin}}, \bibinfo {author} {\bibfnamefont
  {Y.}~\bibnamefont {Meng}}, \bibinfo {author} {\bibfnamefont {G.}~\bibnamefont
  {Shen}}, \bibinfo {author} {\bibfnamefont {L.}~\bibnamefont {Bai}}, \bibinfo
  {author} {\bibfnamefont {Y.}~\bibnamefont {Li}}, \bibinfo {author}
  {\bibfnamefont {J.}~\bibnamefont {Liu}}, \bibinfo {author} {\bibfnamefont
  {H.}~\bibnamefont {kwang Mao}}, \ and\ \bibinfo {author} {\bibfnamefont
  {W.~L.}\ \bibnamefont {Mao}},\ }\href {\doibase
  https://doi.org/10.1016/j.cplett.2009.03.072} {\bibfield  {journal} {\bibinfo
   {journal} {Chem. Phys. Lett.}\ }\textbf {\bibinfo {volume} {473}},\ \bibinfo
  {pages} {72 } (\bibinfo {year} {2009})}\BibitemShut {NoStop}%
\bibitem [{\citenamefont {Chen}\ \emph {et~al.}(2011)\citenamefont {Chen},
  \citenamefont {Zha}, \citenamefont {Chen}, \citenamefont {Shu}, \citenamefont
  {Hemley},\ and\ \citenamefont {Mao}}]{Chen2011}%
  \BibitemOpen
  \bibfield  {author} {\bibinfo {author} {\bibfnamefont {P.-N.}\ \bibnamefont
  {Chen}}, \bibinfo {author} {\bibfnamefont {C.-S.}\ \bibnamefont {Zha}},
  \bibinfo {author} {\bibfnamefont {X.-J.}\ \bibnamefont {Chen}}, \bibinfo
  {author} {\bibfnamefont {J.}~\bibnamefont {Shu}}, \bibinfo {author}
  {\bibfnamefont {R.~J.}\ \bibnamefont {Hemley}}, \ and\ \bibinfo {author}
  {\bibfnamefont {H.-k.}\ \bibnamefont {Mao}},\ }\href {\doibase
  10.1103/PhysRevB.84.104110} {\bibfield  {journal} {\bibinfo  {journal} {Phys.
  Rev. B}\ }\textbf {\bibinfo {volume} {84}},\ \bibinfo {pages} {104110}
  (\bibinfo {year} {2011})}\BibitemShut {NoStop}%
\bibitem [{\citenamefont {Zhu}\ \emph {et~al.}(2012)\citenamefont {Zhu},
  \citenamefont {Oganov}, \citenamefont {Glass},\ and\ \citenamefont
  {Stokes}}]{Zhu2012}%
  \BibitemOpen
  \bibfield  {author} {\bibinfo {author} {\bibfnamefont {Q.}~\bibnamefont
  {Zhu}}, \bibinfo {author} {\bibfnamefont {A.~R.}\ \bibnamefont {Oganov}},
  \bibinfo {author} {\bibfnamefont {C.~W.}\ \bibnamefont {Glass}}, \ and\
  \bibinfo {author} {\bibfnamefont {H.~T.}\ \bibnamefont {Stokes}},\ }\href
  {\doibase 10.1107/S0108768112017466} {\bibfield  {journal} {\bibinfo
  {journal} {Acta Cryst. B}\ }\textbf {\bibinfo {volume} {68}},\ \bibinfo
  {pages} {215} (\bibinfo {year} {2012})}\BibitemShut {NoStop}%
\bibitem [{\citenamefont {Proctor}\ \emph {et~al.}(2017)\citenamefont
  {Proctor}, \citenamefont {Maynard-Casely}, \citenamefont {Hakeem},\ and\
  \citenamefont {Cantiah}}]{Proctor2017}%
  \BibitemOpen
  \bibfield  {author} {\bibinfo {author} {\bibfnamefont {J.}~\bibnamefont
  {Proctor}}, \bibinfo {author} {\bibfnamefont {H.}~\bibnamefont
  {Maynard-Casely}}, \bibinfo {author} {\bibfnamefont {M.}~\bibnamefont
  {Hakeem}}, \ and\ \bibinfo {author} {\bibfnamefont {D.}~\bibnamefont
  {Cantiah}},\ }\href {\doibase 10.1002/jrs.5237} {\bibfield  {journal}
  {\bibinfo  {journal} {J. Raman Spectrosc.}\ }\textbf {\bibinfo {volume}
  {48}},\ \bibinfo {pages} {1777} (\bibinfo {year} {2017})}\BibitemShut
  {NoStop}%
\bibitem [{\citenamefont {Hazen}\ \emph {et~al.}(1980)\citenamefont {Hazen},
  \citenamefont {Mao}, \citenamefont {Finger},\ and\ \citenamefont
  {Bell}}]{Hazen1980}%
  \BibitemOpen
  \bibfield  {author} {\bibinfo {author} {\bibfnamefont {R.~M.}\ \bibnamefont
  {Hazen}}, \bibinfo {author} {\bibfnamefont {H.~K.}\ \bibnamefont {Mao}},
  \bibinfo {author} {\bibfnamefont {L.~W.}\ \bibnamefont {Finger}}, \ and\
  \bibinfo {author} {\bibfnamefont {P.~M.}\ \bibnamefont {Bell}},\ }\href
  {\doibase 10.1063/1.91909} {\bibfield  {journal} {\bibinfo  {journal} {Appl.
  Phys. Lett.}\ }\textbf {\bibinfo {volume} {37}},\ \bibinfo {pages} {288}
  (\bibinfo {year} {1980})}\BibitemShut {NoStop}%
\bibitem [{\citenamefont {Bini}\ and\ \citenamefont
  {Pratesi}(1997)}]{Bini1997}%
  \BibitemOpen
  \bibfield  {author} {\bibinfo {author} {\bibfnamefont {R.}~\bibnamefont
  {Bini}}\ and\ \bibinfo {author} {\bibfnamefont {G.}~\bibnamefont {Pratesi}},\
  }\href {\doibase 10.1103/PhysRevB.55.14800} {\bibfield  {journal} {\bibinfo
  {journal} {Phys. Rev. B}\ }\textbf {\bibinfo {volume} {55}},\ \bibinfo
  {pages} {14800} (\bibinfo {year} {1997})}\BibitemShut {NoStop}%
\bibitem [{\citenamefont {Kurnosov}\ \emph {et~al.}(2006)\citenamefont
  {Kurnosov}, \citenamefont {Ogienko}, \citenamefont {Goryainov}, \citenamefont
  {Larionov}, \citenamefont {Manakov}, \citenamefont {Lihacheva}, \citenamefont
  {Aladko}, \citenamefont {Zhurko}, \citenamefont {Voronin}, \citenamefont
  {Berger},\ and\ \citenamefont {Ancharov}}]{Kurnosov2006}%
  \BibitemOpen
  \bibfield  {author} {\bibinfo {author} {\bibfnamefont {A.~V.}\ \bibnamefont
  {Kurnosov}}, \bibinfo {author} {\bibfnamefont {A.~G.}\ \bibnamefont
  {Ogienko}}, \bibinfo {author} {\bibfnamefont {S.~V.}\ \bibnamefont
  {Goryainov}}, \bibinfo {author} {\bibfnamefont {E.~G.}\ \bibnamefont
  {Larionov}}, \bibinfo {author} {\bibfnamefont {A.~Y.}\ \bibnamefont
  {Manakov}}, \bibinfo {author} {\bibfnamefont {A.~Y.}\ \bibnamefont
  {Lihacheva}}, \bibinfo {author} {\bibfnamefont {E.~Y.}\ \bibnamefont
  {Aladko}}, \bibinfo {author} {\bibfnamefont {F.~V.}\ \bibnamefont {Zhurko}},
  \bibinfo {author} {\bibfnamefont {V.~I.}\ \bibnamefont {Voronin}}, \bibinfo
  {author} {\bibfnamefont {I.~F.}\ \bibnamefont {Berger}}, \ and\ \bibinfo
  {author} {\bibfnamefont {A.~I.}\ \bibnamefont {Ancharov}},\ }\href {\doibase
  10.1021/jp0636726} {\bibfield  {journal} {\bibinfo  {journal} {J. Phys. Chem.
  B}\ }\textbf {\bibinfo {volume} {110}},\ \bibinfo {pages} {21788} (\bibinfo
  {year} {2006})}\BibitemShut {NoStop}%
\bibitem [{\citenamefont {Podsiadło}, \citenamefont {Olejniczak},\ and\
  \citenamefont {Katrusiak}(2017)}]{Podsiadlo2017}%
  \BibitemOpen
  \bibfield  {author} {\bibinfo {author} {\bibfnamefont {M.}~\bibnamefont
  {Podsiadło}}, \bibinfo {author} {\bibfnamefont {A.}~\bibnamefont
  {Olejniczak}}, \ and\ \bibinfo {author} {\bibfnamefont {A.}~\bibnamefont
  {Katrusiak}},\ }\href {\doibase 10.1021/acs.cgd.6b01474} {\bibfield
  {journal} {\bibinfo  {journal} {Cryst. Growth Des.}\ }\textbf {\bibinfo
  {volume} {17}},\ \bibinfo {pages} {228} (\bibinfo {year} {2017})}\BibitemShut
  {NoStop}%
\bibitem [{\citenamefont {Somayazulu}\ \emph {et~al.}(1996)\citenamefont
  {Somayazulu}, \citenamefont {Finger}, \citenamefont {Hemley},\ and\
  \citenamefont {Mao}}]{Somayazulu1996}%
  \BibitemOpen
  \bibfield  {author} {\bibinfo {author} {\bibfnamefont {M.~S.}\ \bibnamefont
  {Somayazulu}}, \bibinfo {author} {\bibfnamefont {L.~W.}\ \bibnamefont
  {Finger}}, \bibinfo {author} {\bibfnamefont {R.~J.}\ \bibnamefont {Hemley}},
  \ and\ \bibinfo {author} {\bibfnamefont {H.~K.}\ \bibnamefont {Mao}},\ }\href
  {\doibase 10.1126/science.271.5254.1400} {\bibfield  {journal} {\bibinfo
  {journal} {Science}\ }\textbf {\bibinfo {volume} {271}},\ \bibinfo {pages}
  {1400} (\bibinfo {year} {1996})}\BibitemShut {NoStop}%
\bibitem [{\citenamefont {Naumova}, \citenamefont {Lepeshkin},\ and\
  \citenamefont {Oganov}(2019)}]{Naumova2019}%
  \BibitemOpen
  \bibfield  {author} {\bibinfo {author} {\bibfnamefont {A.~S.}\ \bibnamefont
  {Naumova}}, \bibinfo {author} {\bibfnamefont {S.~V.}\ \bibnamefont
  {Lepeshkin}}, \ and\ \bibinfo {author} {\bibfnamefont {A.~R.}\ \bibnamefont
  {Oganov}},\ }\href {\doibase 10.1021/acs.jpcc.9b01353} {\bibfield  {journal}
  {\bibinfo  {journal} {J. Phys. Chem. C}\ }\textbf {\bibinfo {volume} {123}},\
  \bibinfo {pages} {20497} (\bibinfo {year} {2019})}\BibitemShut {NoStop}%
\bibitem [{\citenamefont {Spanu}\ \emph {et~al.}(2011)\citenamefont {Spanu},
  \citenamefont {Donadio}, \citenamefont {Hohl}, \citenamefont {Schwegler},\
  and\ \citenamefont {Galli}}]{Spanu2011}%
  \BibitemOpen
  \bibfield  {author} {\bibinfo {author} {\bibfnamefont {L.}~\bibnamefont
  {Spanu}}, \bibinfo {author} {\bibfnamefont {D.}~\bibnamefont {Donadio}},
  \bibinfo {author} {\bibfnamefont {D.}~\bibnamefont {Hohl}}, \bibinfo {author}
  {\bibfnamefont {E.}~\bibnamefont {Schwegler}}, \ and\ \bibinfo {author}
  {\bibfnamefont {G.}~\bibnamefont {Galli}},\ }\href@noop {} {\bibfield
  {journal} {\bibinfo  {journal} {Proc. Natl. Acad. Sci. U.S.A}\ } (\bibinfo
  {year} {2011})}\BibitemShut {NoStop}%
\bibitem [{\citenamefont {Klimenko}, \citenamefont {Gal’tsov},\ and\
  \citenamefont {Prokhvatilov}(2008)}]{Klimenko2008}%
  \BibitemOpen
  \bibfield  {author} {\bibinfo {author} {\bibfnamefont {N.~A.}\ \bibnamefont
  {Klimenko}}, \bibinfo {author} {\bibfnamefont {N.~N.}\ \bibnamefont
  {Gal’tsov}}, \ and\ \bibinfo {author} {\bibfnamefont {A.~I.}\ \bibnamefont
  {Prokhvatilov}},\ }\href {\doibase 10.1063/1.3029759} {\bibfield  {journal}
  {\bibinfo  {journal} {Low Temperature Physics}\ }\textbf {\bibinfo {volume}
  {34}},\ \bibinfo {pages} {1038} (\bibinfo {year} {2008})}\BibitemShut
  {NoStop}%
\bibitem [{\citenamefont {Van~Nes}\ and\ \citenamefont {Vos}(1978)}]{Nes1978}%
  \BibitemOpen
  \bibfield  {author} {\bibinfo {author} {\bibfnamefont {G.~J.~H.}\
  \bibnamefont {Van~Nes}}\ and\ \bibinfo {author} {\bibfnamefont
  {A.}~\bibnamefont {Vos}},\ }\href {\doibase 10.1107/S0567740878007037}
  {\bibfield  {journal} {\bibinfo  {journal} {Acta Cryst. B}\ }\textbf
  {\bibinfo {volume} {34}},\ \bibinfo {pages} {1947} (\bibinfo {year}
  {1978})}\BibitemShut {NoStop}%
\bibitem [{\citenamefont {Syassen}(2008)}]{Syassen2008}%
  \BibitemOpen
  \bibfield  {author} {\bibinfo {author} {\bibfnamefont {K.}~\bibnamefont
  {Syassen}},\ }\href {\doibase 10.1080/08957950802235640} {\bibfield
  {journal} {\bibinfo  {journal} {High Pressure Res.}\ }\textbf {\bibinfo
  {volume} {28}},\ \bibinfo {pages} {75} (\bibinfo {year} {2008})}\BibitemShut
  {NoStop}%
\bibitem [{\citenamefont {Matsui}(2010)}]{Matsui2010}%
  \BibitemOpen
  \bibfield  {author} {\bibinfo {author} {\bibfnamefont {M.}~\bibnamefont
  {Matsui}},\ }\href {\doibase 10.1088/1742-6596/215/1/012197} {\bibfield
  {journal} {\bibinfo  {journal} {J. Phys. Conf. Ser.}\ }\textbf {\bibinfo
  {volume} {215}},\ \bibinfo {pages} {012197} (\bibinfo {year}
  {2010})}\BibitemShut {NoStop}%
\bibitem [{\citenamefont {Prescher}\ and\ \citenamefont
  {Prakapenka}(2015)}]{Prescher2015}%
  \BibitemOpen
  \bibfield  {author} {\bibinfo {author} {\bibfnamefont {C.}~\bibnamefont
  {Prescher}}\ and\ \bibinfo {author} {\bibfnamefont {V.~B.}\ \bibnamefont
  {Prakapenka}},\ }\href@noop {} {\bibfield  {journal} {\bibinfo  {journal}
  {High Pres. Res.}\ }\textbf {\bibinfo {volume} {35}},\ \bibinfo {pages} {223}
  (\bibinfo {year} {2015})}\BibitemShut {NoStop}%
\bibitem [{\citenamefont {Kraus}\ and\ \citenamefont
  {Nolze}(1996)}]{Kraus1996}%
  \BibitemOpen
  \bibfield  {author} {\bibinfo {author} {\bibfnamefont {W.}~\bibnamefont
  {Kraus}}\ and\ \bibinfo {author} {\bibfnamefont {G.}~\bibnamefont {Nolze}},\
  }\href@noop {} {\bibfield  {journal} {\bibinfo  {journal} {J. Appl.
  Crystallogr.}\ }\textbf {\bibinfo {volume} {29}},\ \bibinfo {pages} {301}
  (\bibinfo {year} {1996})}\BibitemShut {NoStop}%
\bibitem [{\citenamefont {Larson}\ and\ \citenamefont
  {Dreele}(2000)}]{Larson2000}%
  \BibitemOpen
  \bibfield  {author} {\bibinfo {author} {\bibfnamefont {A.~C.}\ \bibnamefont
  {Larson}}\ and\ \bibinfo {author} {\bibfnamefont {R.~B.~V.}\ \bibnamefont
  {Dreele}},\ }\href@noop {} {\enquote {\bibinfo {title} {{GSAS}: General
  structure analysis system report {LAUR} 86-748},}\ }\bibinfo {type} {Tech.
  Rep.}\ (\bibinfo  {institution} {Los Alamos National Laboratory},\ \bibinfo
  {year} {2000})\BibitemShut {NoStop}%
\bibitem [{\citenamefont {Boultif}\ and\ \citenamefont
  {Lou\"{e}r}(2004)}]{Boutlif2004}%
  \BibitemOpen
  \bibfield  {author} {\bibinfo {author} {\bibfnamefont {A.}~\bibnamefont
  {Boultif}}\ and\ \bibinfo {author} {\bibfnamefont {D.}~\bibnamefont
  {Lou\"{e}r}},\ }\href@noop {} {\bibfield  {journal} {\bibinfo  {journal} {J.
  Appl. Crystallogr.}\ }\textbf {\bibinfo {volume} {37}},\ \bibinfo {pages}
  {724} (\bibinfo {year} {2004})}\BibitemShut {NoStop}%
\bibitem [{\citenamefont {Glass}, \citenamefont {Oganov},\ and\ \citenamefont
  {Hansen}()}]{Glass2006}%
  \BibitemOpen
  \bibfield  {author} {\bibinfo {author} {\bibfnamefont {C.~W.}\ \bibnamefont
  {Glass}}, \bibinfo {author} {\bibfnamefont {A.~R.}\ \bibnamefont {Oganov}}, \
  and\ \bibinfo {author} {\bibfnamefont {N.}~\bibnamefont {Hansen}},\
  }\href@noop {} {\bibfield  {journal} {\bibinfo  {journal} {Comp. Phys.
  Comm.}\ }\textbf {\bibinfo {volume} {175}},\ \bibinfo {pages}
  {713}}\BibitemShut {NoStop}%
\bibitem [{\citenamefont {Rappe}\ \emph {et~al.}()\citenamefont {Rappe},
  \citenamefont {Casewit}, \citenamefont {Colwell}, \citenamefont {Goddard},\
  and\ \citenamefont {Skiff}}]{Rappe1992}%
  \BibitemOpen
  \bibfield  {author} {\bibinfo {author} {\bibfnamefont {A.~K.}\ \bibnamefont
  {Rappe}}, \bibinfo {author} {\bibfnamefont {C.~J.}\ \bibnamefont {Casewit}},
  \bibinfo {author} {\bibfnamefont {K.~S.}\ \bibnamefont {Colwell}}, \bibinfo
  {author} {\bibfnamefont {W.~A.}\ \bibnamefont {Goddard}}, \ and\ \bibinfo
  {author} {\bibfnamefont {W.~M.}\ \bibnamefont {Skiff}},\ }\href {\doibase
  10.1021/ja00051a040} {\bibfield  {journal} {\bibinfo  {journal} {J. Am. Chem.
  Soc.}\ }\textbf {\bibinfo {volume} {114}},\ \bibinfo {pages}
  {10024}}\BibitemShut {NoStop}%
\bibitem [{\citenamefont {Kresse}\ and\ \citenamefont {Hafner}()}]{Kresse1993}%
  \BibitemOpen
  \bibfield  {author} {\bibinfo {author} {\bibfnamefont {G.}~\bibnamefont
  {Kresse}}\ and\ \bibinfo {author} {\bibfnamefont {J.}~\bibnamefont
  {Hafner}},\ }\href {\doibase 10.1103/PhysRevB.47.558} {\bibfield  {journal}
  {\bibinfo  {journal} {Phys. Rev. B}\ }\textbf {\bibinfo {volume} {47}},\
  \bibinfo {pages} {558}}\BibitemShut {NoStop}%
\bibitem [{\citenamefont {Kresse}\ and\ \citenamefont
  {Furthmuller}(1996)}]{Kresse1996}%
  \BibitemOpen
  \bibfield  {author} {\bibinfo {author} {\bibfnamefont {G.}~\bibnamefont
  {Kresse}}\ and\ \bibinfo {author} {\bibfnamefont {J.}~\bibnamefont
  {Furthmuller}},\ }\href@noop {} {\bibfield  {journal} {\bibinfo  {journal}
  {Comput. Mater. Sci.}\ }\textbf {\bibinfo {volume} {6}},\ \bibinfo {pages}
  {15} (\bibinfo {year} {1996})}\BibitemShut {NoStop}%
\bibitem [{\citenamefont {Kresse}\ and\ \citenamefont
  {Furthm\"uller}()}]{Kresse1996a}%
  \BibitemOpen
  \bibfield  {author} {\bibinfo {author} {\bibfnamefont {G.}~\bibnamefont
  {Kresse}}\ and\ \bibinfo {author} {\bibfnamefont {J.}~\bibnamefont
  {Furthm\"uller}},\ }\href {\doibase 10.1103/PhysRevB.54.11169} {\bibfield
  {journal} {\bibinfo  {journal} {Phys. Rev. B}\ }\textbf {\bibinfo {volume}
  {54}},\ \bibinfo {pages} {11169}}\BibitemShut {NoStop}%
\bibitem [{\citenamefont {Perdew}, \citenamefont {Burke},\ and\ \citenamefont
  {Ernzerhof}(1996)}]{perdew1996}%
  \BibitemOpen
  \bibfield  {author} {\bibinfo {author} {\bibfnamefont {J.~P.}\ \bibnamefont
  {Perdew}}, \bibinfo {author} {\bibfnamefont {K.}~\bibnamefont {Burke}}, \
  and\ \bibinfo {author} {\bibfnamefont {M.}~\bibnamefont {Ernzerhof}},\
  }\href@noop {} {\bibfield  {journal} {\bibinfo  {journal} {Phys. Rev. Lett.}\
  }\textbf {\bibinfo {volume} {77}},\ \bibinfo {pages} {3865} (\bibinfo {year}
  {1996})}\BibitemShut {NoStop}%
\bibitem [{\citenamefont {Grimme}\ \emph {et~al.}()\citenamefont {Grimme},
  \citenamefont {Antony}, \citenamefont {Ehrlich},\ and\ \citenamefont
  {Krieg}}]{Grimme2010}%
  \BibitemOpen
  \bibfield  {author} {\bibinfo {author} {\bibfnamefont {S.}~\bibnamefont
  {Grimme}}, \bibinfo {author} {\bibfnamefont {J.}~\bibnamefont {Antony}},
  \bibinfo {author} {\bibfnamefont {S.}~\bibnamefont {Ehrlich}}, \ and\
  \bibinfo {author} {\bibfnamefont {H.}~\bibnamefont {Krieg}},\ }\href
  {\doibase 10.1063/1.3382344} {\bibfield  {journal} {\bibinfo  {journal} {J.
  Chem. Phy.}\ }\textbf {\bibinfo {volume} {132}},\ \bibinfo {pages}
  {154104}}\BibitemShut {NoStop}%
\bibitem [{\citenamefont {Kresse}\ and\ \citenamefont
  {Joubert}()}]{kresse1999}%
  \BibitemOpen
  \bibfield  {author} {\bibinfo {author} {\bibfnamefont {G.}~\bibnamefont
  {Kresse}}\ and\ \bibinfo {author} {\bibfnamefont {D.}~\bibnamefont
  {Joubert}},\ }\href {\doibase 10.1103/PhysRevB.59.1758} {\bibfield  {journal}
  {\bibinfo  {journal} {Phys. Rev. B}\ }\textbf {\bibinfo {volume} {59}},\
  \bibinfo {pages} {1758}}\BibitemShut {NoStop}%
\bibitem [{\citenamefont {Helvoort}\ \emph {et~al.}(1987)\citenamefont
  {Helvoort}, \citenamefont {Knippers}, \citenamefont {Fantoni},\ and\
  \citenamefont {Stolte}}]{Helvoort1987}%
  \BibitemOpen
  \bibfield  {author} {\bibinfo {author} {\bibfnamefont {K.~V.}\ \bibnamefont
  {Helvoort}}, \bibinfo {author} {\bibfnamefont {W.}~\bibnamefont {Knippers}},
  \bibinfo {author} {\bibfnamefont {R.}~\bibnamefont {Fantoni}}, \ and\
  \bibinfo {author} {\bibfnamefont {S.}~\bibnamefont {Stolte}},\ }\href
  {\doibase https://doi.org/10.1016/0301-0104(87)85092-9} {\bibfield  {journal}
  {\bibinfo  {journal} {Chem. Phys.}\ }\textbf {\bibinfo {volume} {111}},\
  \bibinfo {pages} {445 } (\bibinfo {year} {1987})}\BibitemShut {NoStop}%
\bibitem [{\citenamefont {Birch}(1978)}]{Birch1978a}%
  \BibitemOpen
  \bibfield  {author} {\bibinfo {author} {\bibfnamefont {F.}~\bibnamefont
  {Birch}},\ }\href@noop {} {\bibfield  {journal} {\bibinfo  {journal} {J.
  Geophys. Res.}\ }\textbf {\bibinfo {volume} {83}},\ \bibinfo {pages} {1257}
  (\bibinfo {year} {1978})}\BibitemShut {NoStop}%
\bibitem [{\citenamefont {Bini}\ \emph {et~al.}(1995)\citenamefont {Bini},
  \citenamefont {Ulivi}, \citenamefont {Jodl},\ and\ \citenamefont
  {Salvi}}]{Bini1995}%
  \BibitemOpen
  \bibfield  {author} {\bibinfo {author} {\bibfnamefont {R.}~\bibnamefont
  {Bini}}, \bibinfo {author} {\bibfnamefont {L.}~\bibnamefont {Ulivi}},
  \bibinfo {author} {\bibfnamefont {H.~J.}\ \bibnamefont {Jodl}}, \ and\
  \bibinfo {author} {\bibfnamefont {P.~R.}\ \bibnamefont {Salvi}},\ }\href
  {\doibase 10.1063/1.469810} {\bibfield  {journal} {\bibinfo  {journal} {The
  Journal of Chemical Physics}\ }\textbf {\bibinfo {volume} {103}},\ \bibinfo
  {pages} {1353} (\bibinfo {year} {1995})}\BibitemShut {NoStop}%
\bibitem [{\citenamefont {Loubeyre}\ \emph {et~al.}(1996)\citenamefont
  {Loubeyre}, \citenamefont {LeToullec}, \citenamefont {Hausermann},
  \citenamefont {Hanfland}, \citenamefont {Hemley}, \citenamefont {Mao},\ and\
  \citenamefont {Finger}}]{Loubeyre1996}%
  \BibitemOpen
  \bibfield  {author} {\bibinfo {author} {\bibfnamefont {P.}~\bibnamefont
  {Loubeyre}}, \bibinfo {author} {\bibfnamefont {R.}~\bibnamefont {LeToullec}},
  \bibinfo {author} {\bibfnamefont {D.}~\bibnamefont {Hausermann}}, \bibinfo
  {author} {\bibfnamefont {M.}~\bibnamefont {Hanfland}}, \bibinfo {author}
  {\bibfnamefont {R.~J.}\ \bibnamefont {Hemley}}, \bibinfo {author}
  {\bibfnamefont {H.~K.}\ \bibnamefont {Mao}}, \ and\ \bibinfo {author}
  {\bibfnamefont {L.~W.}\ \bibnamefont {Finger}},\ }\href
  {https://doi.org/10.1038/383702a0} {\bibfield  {journal} {\bibinfo  {journal}
  {Nature}\ }\textbf {\bibinfo {volume} {383}},\ \bibinfo {pages} {702}
  (\bibinfo {year} {1996})}\BibitemShut {NoStop}%
\end{thebibliography}
\end{document}